\documentstyle[12pt,aasms4,epsf]{article}
\received{00 September 1997}
\revised {January 1998}
\accepted{00 1997}

\centerline{Submitted to the Editor of the Astrophysical Journal {\it 
Letters} on November12, 1997}

\lefthead{Yusef-Zadeh, Cotton and Reynolds}
\righthead{A Young SNR Near the Galactic Center\?}

\def\kms{km s$^{-1}$}

\begin{document}

\title{G359.87+0.18: A  Young SNR Candidate Near the Galactic Center?}

\author{F. Yusef-Zadeh}
\affil{Department of Physics and Astronomy, Northwestern University, 
Evanston, Il. 60208 (zadeh@nwu.edu)}

\author{W. D. Cotton}
\affil{National Radio Astronomy Observatory, Charlottesville,
VA 22903 (bcotton@aoc.nrao.edu)}

\author{S. P. Reynolds}
\affil{Physics Department, North Carolina State University,
Box 8202, Raleigh, NC 27695-8202 (steve\_reynolds@ncsu.edu)}

\begin{abstract}

Sub-arcsecond radio continuum observations of the Galactic center
region at $\lambda$6 and 2cm reveal a 0.5$^{\prime\prime}$ diameter
source with a shell-like morphology.  This source is linearly
polarized at a level of 16\% at $\lambda$6cm and has a steep
nonthermal spectrum with spectral index 1.6 between $\lambda$6 and
2 cm. The distance to this source is not known but the large rotation
measure value of 3000 rad m$^{-2}$ suggests that G359.87+0.18 is
likely to be located in the inner Galaxy or at an extragalactic
distance.  
We discuss possible interpretations of this object as a recent
supernova, a very young supernova remnant, 
a nova remnant, or an extragalactic source.  All possibilities
are highly problematic.

\end{abstract}

\keywords{galaxies:  ISM---Galaxy: center ---ISM: individual 
(supernova remnant) --- ISM: magnetic fields}

\vfill\eject

\section{Introduction}

Recent surveys of radio supernova remnants (SNRs) indicate a
underrepresentation of small-diameter young SNRs in the Galaxy (Green
1994).  This deficiency, which is in part due to the lack of
high-resolution observations, is particularly noticeable for SNRs
having diameters less than 1$^\prime$.  There are only four such
sources with diameters less than 4$^\prime$ in the catalogue of Green
(1994).  Because of its large concentration of dense molecular clouds
and massive-star formation, the Galactic center may be considered as a
good target for searching for small diameter SNRs.  There is a
possibility that SNRs in the Galactic center region may be
preferentially smaller in diameter than elsewhere, due to confinement
by a dense interstellar medium in this region (Gray 1994). However, a
search for such objects suffers from considerable confusion due to
bright and extended radio continuum features associated with HII
regions, planetary nebulae, and nonthermal sources in the Galactic
center.

A radio continuum survey of the Galactic center center region in
search of planetary nebulae detected a continuum source, since
named G359.87+0.18, at
$\lambda$20cm using the Westerbork Synthesis Radio Telescope with the
beam size of 22$^{\prime\prime} \times 120^{\prime\prime}$ (Isaacman
1981).  VLA observations of this source were reported with a resolution
of $\approx10^{\prime\prime}$ by Yusef-Zadeh (1986). These
observations were then followed up by Lazio (1997) who studied the
scattering medium toward the Galactic center region and found that the 
source is not heavily scattered. 
Here, we report
high-resolution observations of this radio continuum source,
showing that it is resolved into two components.  The
brighter component is characterized by a nonthermal spectrum,
linear polarization, and shell-like morphology. The sub-arcsecond
diameter of this shell-like polarized source makes it a possible
candidate for a young SNR lying about 15$^\prime$ from the Galactic center
along the rotation axis of the Galaxy.

\section{Observations}

Radio continuum observations of compact radio sources in the Galactic
center were carried out with the Very Large Array of the National Radio
Astronomy Observatory\footnote{The National Radio Astronomy
Observatory is a facility of the National Science Foundation, operated
under a cooperative agreement by Associated Universities, Inc.}  in
its A configuration at $\lambda$2 and 6cm in May 1986.  Each source
was observed for about 5 minutes at each wavelength using 100 MHz of
bandwidth.  A more detailed account of observations and the results of
all the observed compact sources will be given elsewhere. Here we
present the results of one of these compact sources, known as source J
as described by Yusef-Zadeh (1986) and Lazio (1997).  Standard
calibration of all four Stokes parameters was done in AIPS using 3C286 and
1720-130 as the flux and phase calibrators.  The synthesized beam sizes are
0.98$''\times0.34$\arcsec and 0.36$''\times0.13$\arcsec at $\lambda$6
and 2cm, respectively. The $\lambda$20cm continuum emission reported 
here is based on A-array observations described by Yusef-Zadeh et al. 
(1994). The rms noise at this frequency is 1.4 mJy beam$^{-1}$ with a 
beam size of $2.68''\times0.94''$ (PA= --19$^\circ$) Because the phase 
center of this observation is offset considerably from the position of 
the source, bandwidth smearing increases the
source size in the East-West direction to an estimated
0.8$''$.


\section{Results }

Figure 1 shows contours of the total intensity of G359.87+0.18 at
$\lambda$6cm, with rms noise of 0.19 mJy. This source breaks up into two
components A and B with the respective peak flux densities of 22.4 mJy
and 6.5 mJy. Gaussians fitted to these components are at positions
$\alpha (1950)= 17^h 41^m 26^s.45$, $\delta(1950)= -28^0 55' 55.7''$
for source A (l=359.872$^\circ$, b=0.178$^\circ$), and $\alpha (1950)=
17^h 41^m 26^s.21$, $\delta(1950)=-28^0 55' 57.5''$ for source B
(l=359.871$^\circ$, b=0.178$^\circ$).  The brighter source A shows a
5$\sigma$ elongated protrusion with a size of $\approx$0.5$''$ running
in the east-west direction.

 The Gaussians fitted to sources A and B indicate that they are
partially resolved at $\lambda$6cm. This is supported by the
$\lambda$2cm data, with three times higher resolution than the 6cm
data. Figure 2 shows the 2cm image with a resolution of
0.36$''\times0.13''$ and rms noise of 0.34 mJy.  Source A is resolved
into a shell source with a barrel-shaped appearance and a diameter of
about 0.5$''$. No significant polarized emission is detected at
$\lambda$2cm, with an upper limit to the degree of polarization of
28\%.  Source B is also resolved
into an unresolved compact source and a weak extended structure at a
level of 0.8 mJy to the north of the compact source.  The flux density
of the compact component of source B peaks at a level of 2 mJy at the
position of $\alpha (1950)= 17^h 41^m 26^s.21$, $\delta(1950)= -28^0
55' 57.5''$.  More support for the extended nature of these sources
comes from multi-wavelength observations indicating that sources A and
B are not compact and are not affected by the scattering medium toward
the Galactic center region (Lazio 1997).

Figure 3 shows the polarized intensity image of source A at
$\lambda$6cm with rms noise of 0.16 mJy, superimposed on the
distribution of electric field vectors.  Source A has a peak polarized
flux density of 2.1 mJy and appears asymmetric. There are two
polarized clumps having fractional polarization of 16.5\% and 6\% to
the west and to the east, respectively.  The polarized clumps coincide
with the eastern and western edges of the total intensity image as
shown in Figure 1. The true distribution of the magnetic field cannot
be determined with the present data due to large Faraday rotation at
$\lambda$6cm.  The mean rotation measure (RM) toward source A is about
+3000 rad m$^{-2}$, based on two closely spaced frequencies (4860.1
and 4885.1 MHz).  

The east-west sides of the barrel-shape structure noted in the total intensity 
image  of Figure 2 coincide with the clumps of
polarized emission as seen in Figure 3. This indicates that the lack
of polarized emission from the position of the peak of the 6cm total
intensity is not due to depolarization but rather due to the
shell-like morphology of the source as displayed in Figure 2.
Because of the shell-like morphology of the total
intensity at $\lambda$2cm and polarized intensity at $\lambda$6cm, 
the lack of polarized emission between the clumps is
unlikely to be due to the rotation of the plane of polarization across
the synthesized beam or due to internal
Faraday depolarization. We consider it more likely
that the polarized flux shares the shell morphology of source A as
discussed below.

The Gaussian fitted peak flux densities at $\lambda$6 and 2cm with
identical beam size of 0.98$''\times0.32''$ are 22.47
(6.48) and 3.53 (1.52) mJy/beam for source
A(B), respectively. 
Using the rms noise of 0.19 and 0.34 mJy for the $\lambda$6 and 2cm images,
the estimated values of spectral index $\alpha$,
where F$_{\nu} \propto \nu^{-\alpha}$, between $\lambda$2 and 6cm for
sources A and B are 1.6$\pm$ 0.2 and 1.3$\pm$0.5,
respectively. 
The spectral index estimate of 
Source A    is based on using identical  {\it {uv}} coverage between 
50 and 500 $k\lambda$ whereas  that of source B is based on 
slightly different {\it uv} coverage, thus the spectral index 
estimate may be an upper limit 
because some of the
$\lambda$2cm flux may have been resolved out. 
 The spectral index between $\lambda$6 and 20cm,
though uncertain due to unmatched resolutions and to bandwidth
smearing, is about 1.19 when the fluxes from both A and B sources are
added. Source diameter and spectral indices 
 of this source have also been measured 
at a number of frequencies in order to study the characteristics of the 
the scattering screen toward the Galactic center, though with different 
spatial resolutions  (Lazio 1997). 
This again suggests that sources A and B have steep spectral
indices between 2 and 20cm.


\section{Discussion}

A number of studies indicate a ``missing'' population of
small-diameter SNRs in the Galaxy (e.g. Green 1991). Such
small-diameter SNR candidates have been identified in the past but a
majority of them turned out to be thermal sources. The initial
interpretations as SNR candidates were based primarily on shell-like
morphology, a nonthermal interpretation of the high-frequency
spectrum, and a lack of radio recombination line emission (Green 1986,
Cowan et al. 1989, Subrahmanyan, R. 1993; Muizon et al. 1988; Reich et
al. 1984). We believe that the strongest argument that distinguishes
this source from thermal sources such as G25.5+0.2 and G70.68+1.20 is
the evidence for linearly polarized emission from source A of
G359.87+0.18.  Thus even if enough flux is missing from the 2 cm data
to allow a flat spectrum, the interpretation of the emission as
nonthermal seems secure.

However, the source's extremely small size makes detailed
interpretation difficult.  The RM towards source A is about 3000 rad
m$^{-2}$, more than two orders of magnitude greater than RM's toward
sources in the outer Galaxy.  Such rotation measures have been
measured for a number of sources located within a degree of the
Galactic center (Inoue et al. 1984; Yusef-Zadeh \& Morris 1987; Gray
et al.  1995; Yusef-Zadeh, Wardle and Parataran 1996), suggesting that
the Faraday rotation occurs close to or beyond the Galactic center.  A
milliGauss magnetic field strength with an electron density of 0.03
cm$^{-3}$ in the inner 50 pc of the Galactic center can account for
the observed RM (Koyama et al. 1986, 1996; Yamauchi et al. 1990;
Yusef-Zadeh and Morris 1987).  Thus the high RM indicates that
G359.87+0.18 is no closer than the Galactic center; let us initially
presume that it is at the Galactic center, at a distance of 8.5 kpc,
where $1^{\prime\prime} = 0.04$ pc.  Then the source angular radius of
about $0.^{\prime\prime}25$ corresponds to a linear radius of only
0.01 pc.

We first consider an interpretation as a radio SN or SNR.  The mass
contained in this tiny sphere is only $1.6 \times 10^{-7} n_0 \
M_{\odot}$, where $n_0$ is the mean atomic hydrogen density, so unless
G359.87+0.18 is in an exceptionally high-density environment, even for
the GC, it has not swept up an appreciable amount of mass, and should
still be freely expanding.  Then the age is only $10 \ v_8^{-1}$ yr,
where $v_8$ is the ejection velocity in units of 1000 km s$^{-1}$,
expected to be of order 5 for SN Ib or II, and 10 for SN Ia.  A
core-collapse supernova should have emitted a neutrino flux larger
than that of SN1987A by $(55 \ {\rm kpc} / 8.5 \ {\rm kpc})^2$ or
about 42, so might have been seen by any neutrino detectors operating
during the early 1980's.  Furthermore, the mean surface brightness of
G359.87+0.18 Source A is 6200 Jy arcmin$^{-2}$, 45 times higher than
that of Cas A, which is itself far brighter than any other Galactic
SNR (see data in Green 1991).  Thus a supernova-remnant interpretation
is really inappropriate, and we should consider the possibility that
G359.87+0.18 is a radio supernova.

Typical radio supernovae are detected by the VLA with fluxes in the
range of mJy.  For example, van Dyk et al.~(1993) report on five
bright radio supernovae.  All have radio spectral indices of order 1
or less (with errors of order 0.2), much flatter than G359.87+0.18
Source A.  Furthermore, all these have radio fluxes at 6 cm of tens of
mJy, typically, comparable to that of G359.87+0.18, even though they
are farther away by factors of 1000 or more.  Other radio supernovae
(see, e.g., Weiler et al.~1986) are similar in flux density and
spectral index.  So all the radio supernovae we know are more luminous
than G359.87+0.18 by six or more orders of magnitude.  (Of course,
ones as faint as G359.87+0.18 could not be seen in external galaxies;
but the point is that G359.87+0.18 does not resemble known radio
supernovae.)

Could G359.87+0.18 be a nova remnant?  One radio-shell remnant of a
classical nova is known (GK Per; Reynolds and Chevalier 1984, Seaquist et
al.~1989).  But a standard equipartition analysis of G359.87+0.18
(e.g., Pacholczyk 1970) gives a minimum energy in magnetic field and
relativistic electrons of $4 \times 10^{44}$ erg, only a factor of 2
less than the total energy emitted in a classical nova outburst, and
50 times larger than that of GK Per.  So G359.87+0.18 cannot be a nova
remnant.  (The equipartition energy varies with distance $d$ as
$d^{17/7}$, so moving G359.87+0.18 farther away does not help, and the
RM constraint prevents us from moving it closer.)  However, the
equipartition magnetic field one derives is not unreasonable, at about
$6 (d/8.5 \ {\rm kpc})^{17/14} \ \mu$gauss, so that the synchrotron
interpretation of the radio emission is sensible.

Based on different spectra of sources A and B and their morphology, it
is unlikely that these sources are related to each other and therefore
unlikely that G359.87+0.18 is an extragalactic double radio source.
However, it is possible that Source A alone is extragalactic. Its
morphology would be unusual, and its steep spectral index very rare
but not unprecedented. Such ``ultrasteep-spectrum'' radio sources have
been interpreted as radio galaxies at redshifts of order 1. Additional 
problems with the extragalactic interpretation is related to 
the scattering studies 
carried out 
toward this source by Lazio (1997). This author concludes that this source is 
unlikely to be extragalactic and affected by a scattering screen seen 
toward the Galactic center (e.g. van Langevelde et al. 1992).
This conclusion is based on the predicted value of the observed diameter 
of this source at 0.33 GHz being  25\% larger than the observed value.
The predicted source diameter  
assumes minimal scattering toward this source 
and an intrinsic source size $\theta_i$=0 (Lazio 1997).

We are forced to conclude that G359.87+0.18 is not well explained
by any known class of nonthermal radio source.  Further observations
are clearly called for to resolve its nature.

\acknowledgments

F. Yusef-Zadeh's work was supported in part by NASA grant NAGW-2518.
S.~Reynolds acknowledges support from NASA grants NAG 5-2844. We thank
Dale Frail for useful discussion.

 
\vfill\eject
\clearpage

\begin{figure}
\plotone{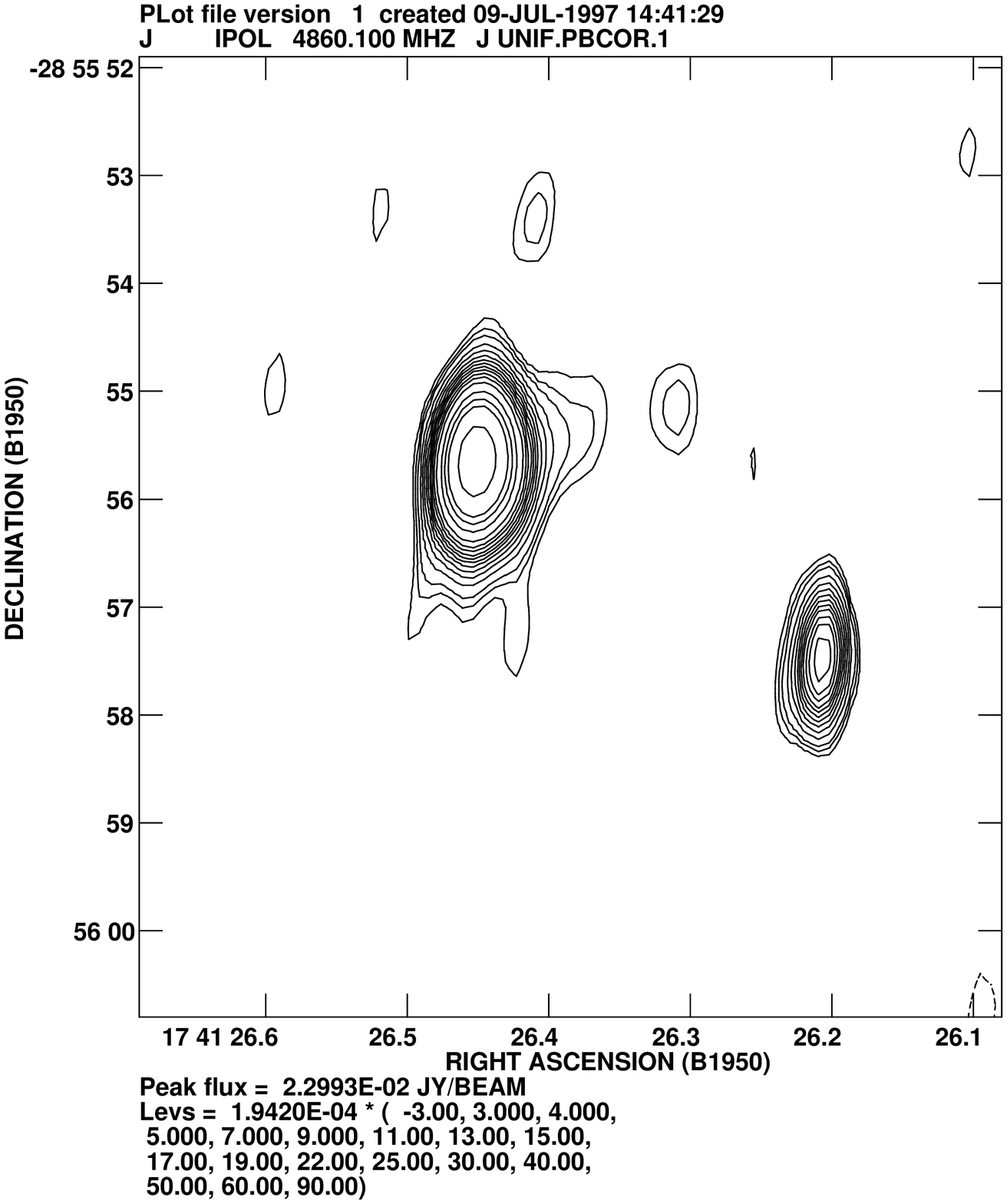} 
\figcaption {Contours of total intensity showing sources A (NE) and B (SW)
with levels set 
at $\lambda$6cm set (-3, 3, 4, 5, 7, 9, 11, 13, 15, 17, 19, 22, 25, 30,
40, 50, 60, 90)$\times$ the rms noise which is 0.194 mJy beam$^{-1}$.
 The spatial resolution 
is 0.98$''\times$0.34$''$ (PA=-3.9$^0$).}
\end{figure}

\begin{figure}
\plotone{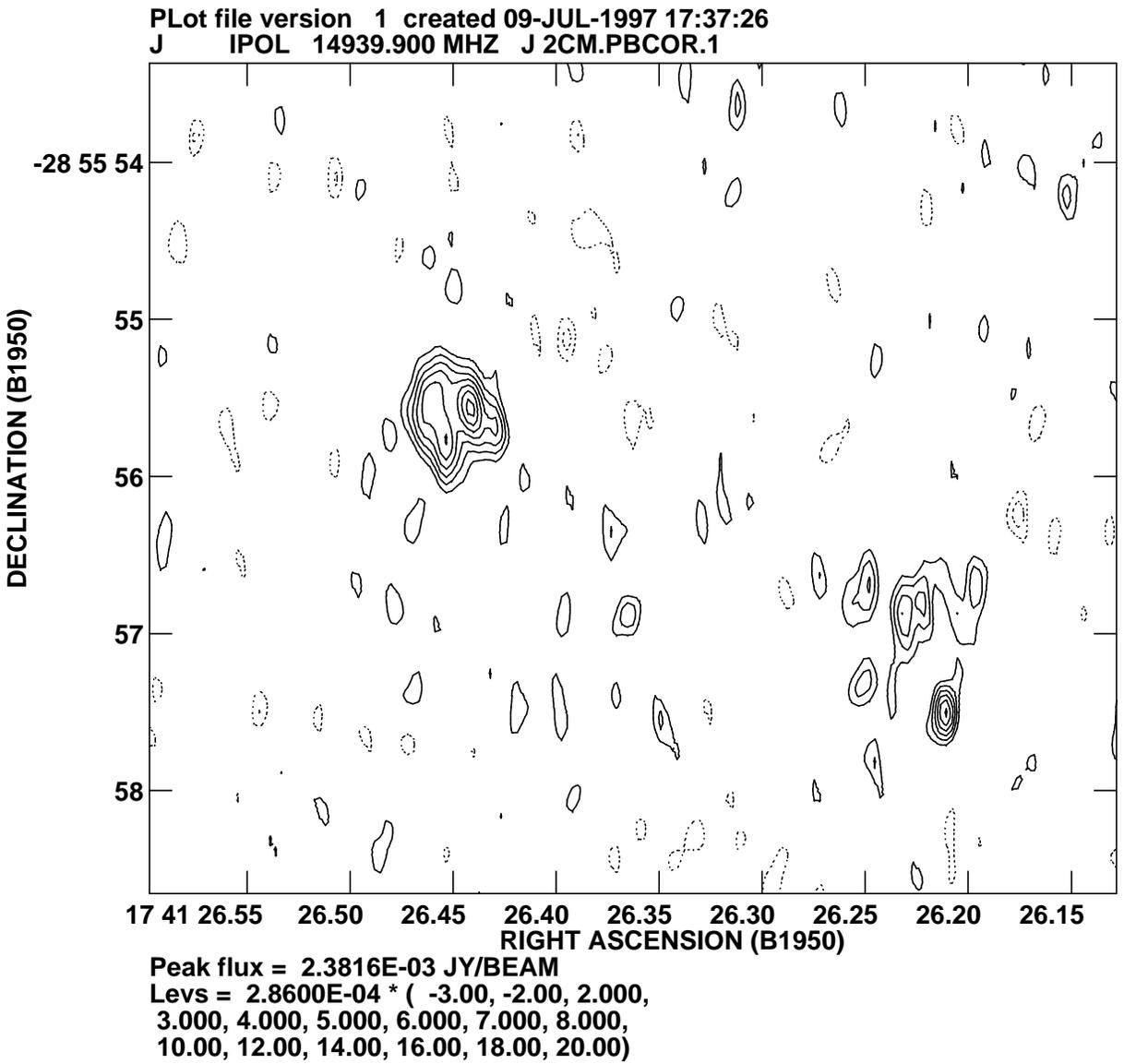}
\figcaption {Contours of total intensity showing sources A  and B 
at $\lambda$2cm with levels set at (-3, -2, 2, 3, 4, 5, 6, 7, 8, 10)
$\times$ the rms noise which is 0.286 mJy beam$^{-1}$.
 The spatial resolution 
is 0.36$''\times$0.13$''$ (PA=6.8$^0$).}
\end{figure}

\begin{figure}
\plotone{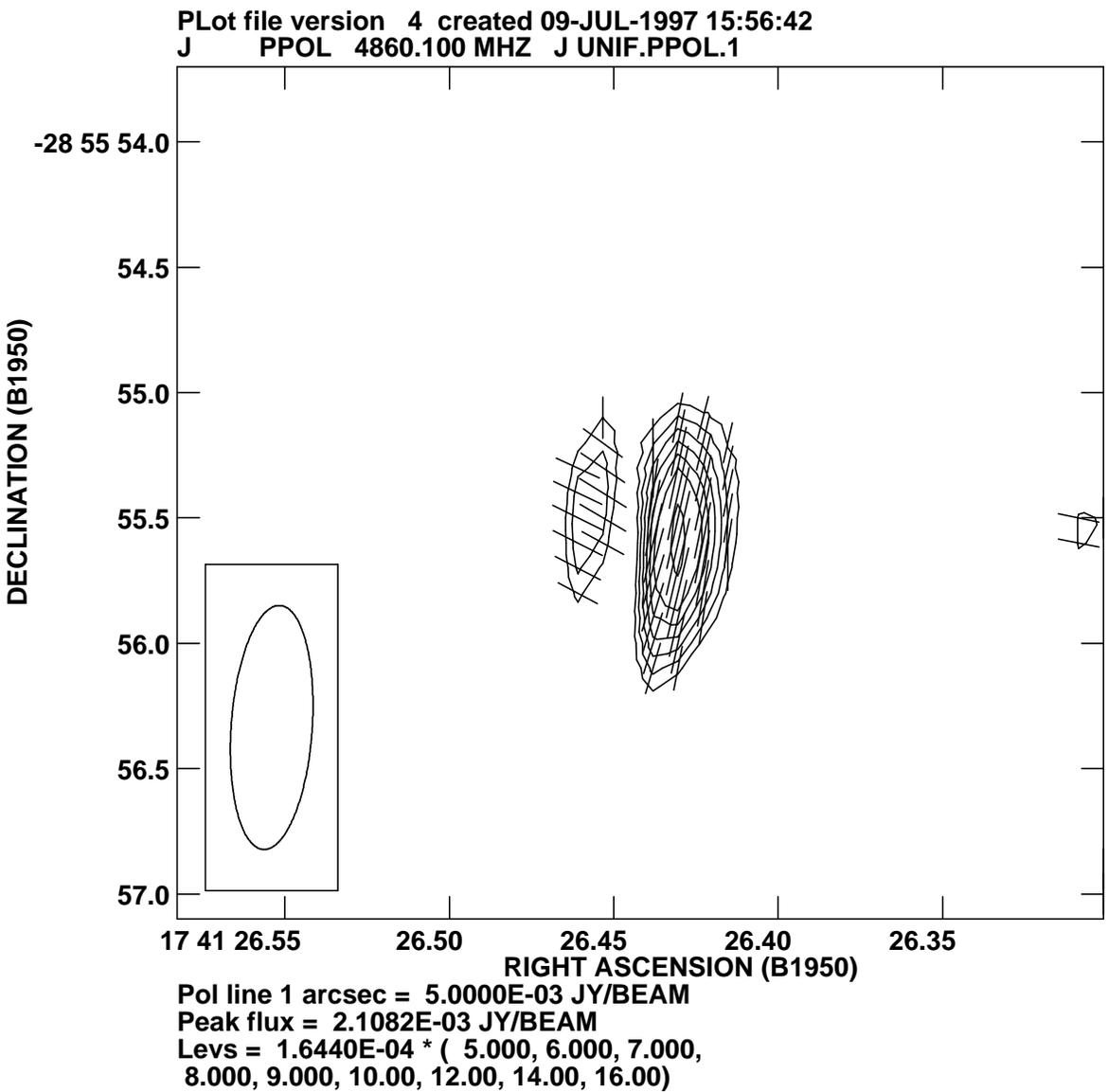}
\figcaption {Contours of polarized  intensity of source A 
at $\lambda$6cm with levels set at 
(5, 6, 7, 8, 9, 10, 12) $\times$ the rms noise 
0.164 mJy beam$^{-1}$ are superimposed on the 
distribution of electric field vectors represented by straight lines. 
The lengths of the straight lines are fixed.
and  the maximum fractional polarization is 16\%.   
 The spatial resolution is identical to that of Figure 1.} 
\end{figure}

\end{document}